\documentstyle[12pt]{article}
\begin{document}
{\it Cavendish Preprint HEP 93/2 \\
16th May 1993}
\vspace{1.0cm}
\begin{center}
{\huge \bf Gauge symmetry and the EMC spin effect \\}
\vspace{3ex}
\vspace{3ex}
{\large \bf S. D. Bass \\
\vspace{3ex}
{\it HEP Group, Cavendish Laboratory, \\
University of Cambridge, Madingley Road, Cambridge, CB3 0HE,
England \\}
\vspace{3ex}                            }
\vspace{3ex}
{\large \bf Abstract \\}
\end{center}
\vspace{3ex}
{\large
We emphasise the EMC spin effect as a problem of symmetry and
discuss the renormalisation of the
$C=+1$ axial tensor operators. This involves the generalisation
of the Adler-Bell-Jackiw anomaly to each of these operators.
We find that the contribution of the axial anomaly to the spin
dependent structure function $g_1 (x, Q^2)$ scales at $O(\alpha_s)$.
This means that the anomaly can be a large $x$ effect in $g_1$.
Finally we discuss the jet signature of the anomaly.}

\pagebreak

\section {\bf Introduction}

There has recently been a great deal of interest in polarised
deep inelastic scattering as a result of the European Muon Collaboration
(EMC [1]) measurement of the spin dependent proton structure
function $g_1^P(x,Q^2)$.
In the naive parton model $g_1$ is written as :
$$
g_{1}(x) = {1 \over 2} \sum_{q} {\rm e_{q}}^{2} \Delta q(x)
\eqno(1)
$$
where
$$
\Delta q(x) = (q^{\uparrow} + {\overline q}^{\uparrow})(x) -
			(q^{\downarrow} + {\overline q}^{\downarrow})(x)
\eqno(2)
$$
is the polarised quark distribution.
We write $\Delta q = \int_0^1 dx \Delta q(x)$.
Then the flavour singlet contribution to the first moment of $g_1$
is the
fraction of the proton's spin that is carried by its quarks in this
parton model.
The EMC data was combined with an earlier measurement of $g_1$
from SLAC [2].
After a smooth Regge extrapolation of their data ($g_1 \sim x^{-0.12}$)
EMC determined the ``spin content" to be [1]
$$
\Delta u + \Delta d + \Delta s = 0.120 \pm 0.094 (stat.) \pm 0.138 (syst.)
\eqno(3)
$$
which is consistent with zero and two standard deviations
from the Ellis-Jaffe hypothesis, which says that strange quarks should
not play a significant role.

There have been many suggestions of how to reconcile this EMC spin effect
with quark model expectations.
(The EMC experiment and its theoretical interpretation are discussed
in detail in the reviews [3] and [4].)
As a model independent statement, we can say that the spin effect is a
violation of OZI in the flavour singlet channel [5].
In QCD the proton ``spin content" (equ.(3)) is the flavour singlet
axial charge $g_A^0$; it is the proton matrix element of the flavour singlet
axial-vector current.
Here we encounter the axial anomaly, which is an effect in the renormalised
axial-vector current operator.
(It is not present in the classical version of this operator.)
The breakdown of the Ellis-Jaffe hypothesis for $g_A^0$ has led to the
EMC spin effect being thought of as a first moment problem.

In truth the axial anomaly affects more than an just one operator.
In the previous paper (ref. [6]) it was shown that the axial anomaly
is a property of
{\it each} moment of $g_1$ and not just the first moment;
that is, the anomaly is relevant to the {\it whole structure function
$g_1 (x, Q^2)$}.
This result involves the generalisation of the axial anomaly to each of the
higher spin axial-tensor operators
${\overline q} \gamma_{\mu} \gamma_5 (iD_{\mu_1})...(iD_{\mu_{2n}})q$.

In this paper we develop these ideas further.
We discuss the model independent properties of the anomaly contribution
to $g_1$ that can be deduced from QCD alone.
Thus we derive constraints
that any future models of proton structure which include the anomaly
will have to satisfy.
Through-out this paper we develop the idea that the EMC spin effect
is a problem in symmetry
and emphasise that we have to be very careful to understand exactly
what we
mean by a parton in QCD.

In Sects.2 and 3 we focus on the role of the anomaly in the inclusive
DIS cross section and work within the framework of the
operator product expansion.
In Sect.2 we explain precisely what
it means to say that the anomaly is relevant to $g_1(x,Q^2)$.
This allows us understand the subtle relationship between spin and
$g_1$.
We show that the anomaly couples in a pointlike (local) manner to
the hard photon in DIS.
In Sect.3 we derive the renormalisation group properties of the anomaly
in $g_1(x,Q^2)$
and show that it scales at $O(\alpha_s)$.
These results suggest that the anomaly may be a large $x$ effect in $g_1$.
An anomaly contribution at
large $x$ would be seen as an OZI breaking effect in the large $x$ bins of
the polarised DIS experiment.
It would show up as a finite difference between $g_1$ and the C-odd spin
dependent proton structure function $g_3$ [7],
which is anomaly free and has the naive parton model interpretation
$(q-{\overline q})^{\uparrow}(x)-(q-{\overline q})^{\downarrow}(x)$.
In Sect.4 we discuss the jet signature of the anomaly.

\section {\bf Spin and $g_1$}

In QCD the deep inelastic process is described by the operator
product expansion (OPE) and the renormalisation group.
The interesting physics of $g_1$ is in the flavour singlet
part, which receives contributions
from both quark and gluon partons, viz.
$$
g_{1}(x, Q^{2}) |_{S} = {1 \over 3} \sqrt {2 \over 3}
\int_x^1 {dz \over z}
\Biggl[ \Delta q_{0}(z, Q^2)
C^{q}_{S} ( {x \over z}, \alpha_s (Q^{2}) )
+ {1 \over \sqrt{6} } \Delta g(z, Q^2)
C^g_S ({x \over z}, \alpha_s (Q^2) ) \Biggr]
\eqno(4)
$$
The C-even, spin dependent quark $\Delta q_0(x, Q^2)$
and gluon $\Delta g(x, Q^2)$ distributions contain all the information
about the target.
They are defined with
respect to the operator product expansion so that their odd moments
project out the target matrix elements
of the renormalised spin odd, composite operators
$$
2M s_+ (p_+)^{2n} \int^1_0 dx \ x^{2n} \Delta q_{k} (x, Q^2) =
<p,s | [ {\overline q}(0) \gamma_+ \gamma_5 (i D_+)^{2n} {\lambda^k \over 2}
q(0) ]_{Q^2}^{GI} |p,s >_c
\eqno(5)
$$
$$
2M s_+ (p_+)^{2n} \int^1_0 dx \ x^{2n} \Delta g (x, Q^2) =
<p,s | [ {\bf \rm Tr} \ G_{+ \alpha}(0) (iD_+)^{2n-1}
{\tilde G}^{\alpha}_{\ +}(0) ]_{Q^2}^{GI} |p,s >_c
(n \geq 1)
\eqno(6)
$$
Here $G_{\mu \nu}$ is the gluon field tensor,
$ {\tilde G}_{\mu \nu} = {1 \over 2} \epsilon_{\mu \nu \alpha \beta}
G^{\alpha \beta} $ is the corresponding dual tensor and
$D_{\mu} = \partial_{\mu} + ig A_{\mu} $ is the gauge covariant
derivative in QCD. The kinematic variables $p_{\mu}$ and $s_{\mu}$
denote the proton momentum and spin vectors respectively ($M$ is the
proton mass).
The superscript on the operators emphasises that they are quoted with
respect to some gauge invariant regularisation scheme,
and the
subscript $Q^2$ indicates the subtraction point.
(We have chosen to write the composite operators at the
renormalisation scale $\mu^2=Q^2$.)
The association of $\Delta q_k(x, Q^2)$ with quarks and $\Delta g(x, Q^2)$
with gluons follows
when we evaluate the target matrix elements in equs.(5-6) in the light-cone
gauge, where $D_+ \rightarrow \partial_+$.

The Wilson coefficient distributions $C(x,\alpha_s)$ are determined by the
leading light-cone singularity in the product of the two electromagnetic
currents on the light cone.
They are target independent and describe the
interaction of the hard photon $\gamma^{*}$ with each of the quark and
gluon partons, which are defined by equs.(5-6).
The coefficients may be calculated
in perturbation theory, where one finds
$C^q(x,\alpha_s)=\delta (x-1)+O(\alpha_s)$ and $C^g(x,\alpha_s)\sim \alpha_s$.

It is known from unpolarised DIS experiments that the gluon distribution
is concentrated at small $x$. In polarised DIS
the hard photon scatters from a gluon via a quark-antiquark
pair, described in $C^g (x, \alpha_s)$.
This dissipates the gluon's already small momentum so that
$\Delta g(x, Q^2)$ is relevant to $g_1$ only at small $x$ ($x \leq 0.03$)
[8]. It makes a negligible contribution to the measured
sum rule between $x = 0.01$ and 1, where the three constituent quarks are
expected to dominate.
The clue to understanding the spin effect lies in the identification
of the
axial-vector current (and the higher spin axial tensors in equ. (5))
with spin.
We need to understand how spin is realised in renormalised QCD.

The parton distributions are renormalised quantities which should always
be quoted with respect to some regularisation.
In general the introduction of a regulator breaks some symmetry of
the classical theory and this symmetry is not always restored in the
continuum limit.
(For a complete discussion of regulator theory we refer to the text of
Zinn-Justin [9] and the lectures of Guichon [10].)
This means that the symmetry properties of the {\it physical}, renormalised
world in which we live are not identical to those of the
classical Lagrangian.
Noether's theorem which tells us about the symmetry of a classical
Lagrangian may not be valid in the renormalised world.
Using a different choice of regulator corresponds to retaining different
symmetries after renormalisation so that
two different
renormalisation procedures can lead to worlds with different symmetry
properties. This clash of symmetries is what is meant by an anomaly.
The symmetries that must be preserved at all costs are Lorentz and gauge
invariance.

It is well known that we cannot choose both gauge invariance and also
to have
chirality as a good quantum number (the Adler-Bell-Jackiw axial
anomaly [11,12]).
The axial anomaly predicts the decay rate for $\pi^0 \rightarrow 2\gamma$.
This decay would be strongly suppressed without the anomaly.
The axial anomaly is also relevant to our understanding of the U(1)
problem in QCD [13] and the EMC proton spin data.
It is important to realise that the anomaly is not restricted to an effect
in the axial-vector current.
An anomaly will manifest itself
where-ever the quantum numbers of the relevent operator allow it.
For example,
any renormalisation calculation involves a subtraction scale - and this
breaks the scale invariance of the classical theory.
The only operators which are scale invariant are those which are not
renormalised (ie. conserved currents and the other ``soft" operators
- the divergences of which just involve mass terms).
The QCD evolution of the parton densities with increasing $Q^2$ and
asymptotic freedom
are intimately related to the scale anomaly.

We now consider the relationship between the axial anomaly and spin.
Classically the axial vector current looks like a gauge invariant
operator.
The quark field operator transforms as
$$
q(x) \rightarrow U(x) q(x)
$$
and
$$
{\overline {q}}(x) \gamma_{\mu} \gamma_{5} \rightarrow
  {\overline {q}}(x) \gamma_{\mu} \gamma_{5} U^{\dagger}(x)
\eqno(7)
$$
under a given gauge transformation $U$.
On the other hand, in quantum field theory
the axial vector current operator is not just
${\overline q}(0) \gamma_{\mu} \gamma_{5}$
multiplied by $q(0)$. It is a composite operator which has to be
renormalised and there are extra
divergences which are intrinsic to the operator itself.
It turns out that one cannot renormalise the axial tensor operators
in a gauge invariant way so that they describe spin at the same time.
In general,
for a given choice of renormalisation prescription $R$, the renormalised
axial tensor operator
differs from the gauge invariant operator by a multiple of a
gauge-dependent, gluonic counterterm $k_{\mu \mu_1 ... \mu_{2n}}$, viz.
$$
\biggl[
{\overline q}(0) \gamma_{\mu} \gamma_5 D_{\mu_1} ... D_{\mu_{2n}} q(0)
\biggr]^R_{Q^2} =
\biggl[
{\overline q}(0) \gamma_{\mu} \gamma_5 D_{\mu_1} ... D_{\mu_{2n}} q(0)
\biggr]^{GI}_{Q^2}
+ \lambda_{R, n} \biggl[ k_{\mu \mu_1 ... \mu_{2n} } \biggr]_{Q^2}
\eqno(8)
$$
where the coefficients $\lambda_{R, n}$ are fixed by the choice of
renormalisation prescription.
This axial anomaly was discovered for the axial vector current in QED.
The operator product expansion is a rigorous result
which follows from the completeness of gauge-invariant local operators
- it is also describing a physical process.
Thus we have to be careful to ensure that our parton distributions are
quoted with respect to some gauge-invariant renormalisation prescription.
(The scheme dependence of parton distributions is a statement about
varying {\it within} the class
of gauge invariant renormalisation prescriptions.)

Not only does the axial anomaly lead to a difference between the
renormalised axial currents which preserve gauge invariance and chiral
symmetry, but in addition the gauge invariant axial current is scale
dependent (in this case the scale is $Q^2$). (The anomalous dimension of
the first moment, $\Delta q_{0}$,was first calculated for QCD by Kodaira
[14].)
This means that one cannot derive the generators of the spin algebra
SU(2) from it [4].
Thus, the gauge invariant axial-vector current does not measure quark
spin in the proton.
The axial-vector current which should be identified with spin
is the renormalisation group invariant symmetry current
[15], viz.
$$
j_{5 \ \mu}^S = j_{5 \ \mu}^{GI} - k_{\mu},
\eqno(9) $$
which satisfies the divergence equation
$\partial^{\mu} j_{5 \mu}^S = 2 m_q {\overline q} i \gamma_5 q$.
Here
$$
k_{\mu} = {\alpha_s \over 2 \pi}
\epsilon_{\mu \lambda \alpha \beta} {\bf Tr} \ A^{\lambda}
(G^{\alpha \beta}-{2 \over 3} A^{\alpha} A^{\beta})
\eqno(10)
$$
is the Chern-Simons current.
The symmetry current $j^S_{5 \mu}$ is {\it the} axial-vector current
in a world where chirality is
an exact symmetry and where gauge invariance is not.
One can say that {\it the gauge symmetry screens the spin of the quarks.}

Going beyond the axial-vector current,
the distribution which does measure quark spin in the proton is defined
with respect to the axial-tensor operators {\it in this same
world} where chirality is an exact symmetry.
In other words,
the spin distribution is constructed formally using operators
which are defined with respect to the same
renormalisation prescription as the symmetry current $j^S_{5 \mu}$.
These S-prescription operators correspond
to particular values of the coefficients $\lambda_n$ in equ.(8),
which were derived in ref.[6]. (These calculations
were done in light-cone gauge $A_+=0$ using dimensional regularisation,
where the anomaly becomes a problem
of how to continue $\gamma_5$ into the regulator dimensions [16].)
One finds
$$
\lambda_{S,n} = - {1 \over (2n+1) (2n+2)}
\eqno(11)
$$
and the two gluon part of $k_{\mu \mu_1 ... \mu_{2n}}$
is
$$
{\alpha_s \over \pi} i^{2n}
\epsilon_{\mu \lambda \alpha \beta} A^{\alpha}_a \partial^{\lambda}
\partial_{\mu_1} ... \partial_{\mu_{2n}} A^{\beta}_a.
\eqno(12)
$$
The spin distribution differs from the
physical distribution $\Delta q (x, Q^2)$ measured in deep inelastic
scattering by an anomalous
gluonic term related to the $k_{\mu \mu_1 ... \mu_{2n}}$.
This clash of symmetry between spin and gauge invariance defines the
anomaly in $g_1$.

We now illustrate this discussion by comparing $g_1$ with the other
structure functions which are measured in DIS experiments.
The axial anomaly is not relevant to the unpolarised quark distributions,
which are described in OPE language by the operators
$ {\overline q}(0) \gamma_+ (iD_+)^{n} q(0)$.
It is also not relevant to the C-odd spin dependent structure function
$g_3$, which is the polarised version of $F_3$.
(In the naive parton model $g_3$ measures
$(q-{\overline q})^{\uparrow}(x) - (q-{\overline q})^{\downarrow}(x)$.)
Since $g_3$ is odd under charge conjugation,
it can have no anomalous gluonic contribution.
(Gluons are C-even.)
This means that it does make sense to talk about $F_1$, $F_3$ and $g_3$
in terms of quarks with explicit spin degrees of freedom - the clash of
symmetry between gauge invariance and spin does not manifest itself in
these structure functions.
We write the gauge invariant distribution as
$$
q^{\uparrow}_{GI}(x, Q^2) = ( q^{\uparrow}_S + {1 \over 4} \kappa ) (x, Q^2)
\eqno(13a)
$$
$$
q^{\downarrow}_{GI}(x, Q^2) = ( q^{\downarrow}_S - {1 \over 4} \kappa ) (x,
Q^2)
\eqno(13b)
$$
for both quarks and anti-quarks.
Here $\kappa (x, Q^2)$ denotes the anomaly and the subscript $S$ denotes
the gauge-dependent spin distribution.
The $\kappa$ part of the quark distribution is relevant only to $g_1$ in
deep inelastic scattering.
It is constructed formally as follows.
We consider the matrix element of the gauge dependent counterterm
$k_{\mu \mu_1...\mu_{2n}}$ acting between nucleon states away from the
forward direction.
This matrix element will contain several form
factors -
some of which are sensitive to the Kogut-Susskind pole,
which features in the solution of the U(1) problem [13].
Here we only need
to consider the axial form factor which
multiplies the Lorentz structure $s_{\mu} (p+p')_{\mu_1}...(p+p')_{\mu_2n}$,
symmetrised over all indices.
(We denote this form factor $\kappa^{(2n)}_A((p-p')^2)$;
$p_{\mu}$ and $p'_{\mu}$ are the nucleon's initial and final momentum.)
We take the forward limit $p'_{\mu} \rightarrow p_{\mu}$ and obtain
a contribution $\kappa^{(2n)}_A(0)$ to the OPE matrix elements in
equ.(5).
The Mellin theorem is applied to these $\kappa_A^{(2n)}(0)$ to obtain
the distribution $\kappa (x, Q^2)$.
It is the difference between the measured quark distribution and the
unphysical distribution which measures spin.

If we work in light-cone gauge then it follows that the three gluon part
of $k_+$ (see eq.(10)) vanishes.
Indeed, it was shown in ref.[17] that the gauge dependent operator
$k_{\mu}$
coincides with a gauge invariant but non-local string operator
in the axial gauge.
It seems reasonable to postulate that $\kappa(x,Q^2)$ is invariant
under gauge transformations within the axial gauge.
In more general (eg. covariant) gauges one can show [15, 18] that
$k_{\mu}$ is not invariant under gauge transformations.
One should not try to isolate the nucleon matrix element of $k_{\mu}$
(or $\kappa(x,Q^2)$) from $\Delta q_0(x,Q^2)$.

The anomaly $\kappa(x,Q^2)$
describes part of the quark Fock state wavefunction of the proton
where
the quarks carry infinite momenta [19, 20].
For example, when the axial tensor couples to the proton via a gluonic
intermediate state (Fig. 1) the anomaly is the purely pointlike part
of the Adler-Bell-Jackiw triangle -
hence the gluonic counterterm $k_{\mu \mu_1 ... \mu_{2n}}$.
Since the anomaly is independent of quark flavour the same $\kappa$
term is relevant to $u, d, s, c, ..$
The anomaly $\kappa$ varies only according to the ``spin" and not the charge
or flavour.
Because the anomaly is flavour singlet
it will induce some OZI violation where-ever it plays a role.
This is the source of the EMC spin effect.

In DIS the physical distributions
$q_{GI}^{\uparrow}(x,Q^2)$ and $q_{GI}^{\downarrow}(x,Q^2)$
couple locally
to the hard photon via the OPE quark coefficient $C^q(x,\alpha_s)$.
When we
substitute equs.(13) into equ.(4) it follows that each of the spin
and $\kappa$ components of the measured quark distribution have the
same pointlike coupling to the photon.
The anomaly couples to the photon in exactly the same way as a quark.
Physically, this means that there is a new local interaction between the
hard photon and a gluonic component in the proton
which must be included into the QCD improved parton model
(see Sect.4).
This is despite the fact that the glue does not carry electric charge !
One can apply this
argument in reverse to say that $C^q(x, \alpha_s)$ defines the
OPE spin dependent quark distribution to include all partons which make
a local interaction with the hard photon in DIS [21].
Clearly, the anomaly is an entirely different gluon effect in $g_1$
than $\Delta g(x, Q^2)$.
We now discuss the renormalisation group properties of the anomaly
in $g_1(x, Q^2)$.

\section {Renormalisation group properties of the anomaly in $g_1$}

At $O( \alpha_s )$ the measured parton distributions evolve with
increasing $Q^2$
according to the GLDAP equations [22]
$$
{d \over dt} \Delta q_0 (x, t) = {\alpha_s \over 2\pi }
\int_x^1 {dz \over z} \Biggl[ \Delta q_0 (z, t) P^q_{qq} ({x \over z}) +
2 N_f \Delta g (z, t) P^q_{gg} ({x \over z}) \Biggr]
$$
$$
{d \over dt} \Delta g (x, t) = {\alpha_s \over 2\pi }
\int_x^1 {dz \over z} \Biggl[ \Delta q_0 (z, t) P^g_{qq} ({x \over z}) +
\Delta g (z, t) P^g_{gg} ({x \over z}) \Biggr]
\eqno(14)
$$
where $t = \ln Q^2$.
The splitting functions have a probability interpretation in the light-cone
gauge so that eg.
$P^q_{gg}(x)$ is the probability
for a gluon to branch into a quark-antiquark pair so that the quark
(or antiquark) carries a fraction $x$ of the gluon's momentum in the
collinear direction.
Formally, the GLDAP equations are related to the OPE in that the moments
of the splitting functions reproduce the anomalous dimensions of the
operators which appear in the OPE.
It is important to emphasise that the spin dependent quark distribution
which appears in the GLDAP equations is the gauge invariant one.
This follows because
the twist-two, gauge-invariant operators which define the parton
distributions
$\Delta q_{0}(x,Q^2)$ and $\Delta g(x,Q^2)$ mix under renormalisation
independently of any gauge non-invariant operators [23].

We wish to disentangle an evolution equation for $\kappa$ from the
GLDAP equations.
To do this we need to evaluate the anomalous dimension of the
counterterm $k_{\mu \mu_1 ... \mu_{2n}}$, which is determined from
its matrix elements
between external quark and gluon states.
In perturbation theory, the anomaly is a purely pointlike part
of the
one-loop vector (gluon), vector (gluon), axial-tensor amplitude
(Fig.2).
Hence there is one power of $\alpha_s$ which is intrinsic to
$k_{\mu \mu_1 ... \mu_{2n}}$ (see equ.(12)).
The
anomalous dimension of $k_{\mu \mu_1...\mu_{2n}}$ starts at
$O(\alpha_s^2)$; it is a two loop effect.
Thus we deduce that the anomaly scales at $O(\alpha_s)$, viz.
$$
{d \over dt} \kappa (x,t) = 0.
\eqno(15a)
$$
All of the QCD evolution of $\Delta q(x, Q^2)$ at $O(\alpha_s)$
is carried
by its spin component.
If we substitute equs.(13) into equs.(14) we can re-write the GLDAP
equations as
$$
{d \over dt} \Delta g(x, t) =
{\alpha_s \over 2 \pi} \int_x^1 {dz \over z}
\Biggl[ ( \Delta q_S (z, t) + \kappa (z, t) )
P^g_{qq} ({x \over z}) + \Delta g(z, t) P^g_{gg} ({x \over z}) \Biggr]
\eqno(15b)
$$
$$
{d \over dt} \Delta q_S(x, t) = {\alpha_s \over 2 \pi}
\int_x^1 {dz \over z} \Biggl[ (\Delta q_S (z, t) + \kappa(z, t) ) P^q_{qq} ({x
\over z}) +
2 N_f \Delta g(z, t) P^q_{gg} ({x \over z}) \Biggr]
\eqno(15c)
$$
Since the physical distribution $\Delta q_0 (x,t)$ evolves to zero at large
$x$ in the asymptotic limit $Q^2 \rightarrow \infty$
it follows that
the spin component must evolve to $ - \kappa (x)$ at infinite $Q^2$ and
finite $x$.

The two loop splitting function corresponding to the anomaly can be
derived as follows.
The spin dependent distribution $\Delta q^{(g)}(x,Q^2)$
of a gluon is evaluated in perturbation theory by calculating the
graph in Fig.2, where we work in the light-cone gauge $A_+=0$.
This calculation has been carried out in refs.[20, 24, 6].
There is a contribution from finite $k_T^2$ between the quark mass
squared and $Q^2$, which has the usual partonic Fock state interpretation,
and
also the anomaly.
The anomaly is a purely pointlike part of the triangle amplitude;
it makes a contribution
$- {\alpha_s \over \pi} N_f (1-x)$ to
$\Delta q^{(g)}(x, Q^2)$ [6], where $N_f$ is the number of quark flavours.
The two loop splitting function for $q \rightarrow \kappa$ is
given by folding this factor with the splitting function for
$q \rightarrow g$, viz.
$P^g_{qq} (x) = {4 \over 3} (2-x)$.
We find
$$
P^{\kappa}_{qq} (x, \alpha_s) = - {4 \alpha_s \over 3 \pi} N_f
\int_x^1 {dz \over z} (1-z) (2 - {x \over z})
\eqno(16)
$$
Equ.(16) is the generalisation of Kodaira's two loop anomalous dimension
of the axial vector current to the higher spin axial tensor operators.
Kodaira's result [14] can be obtained
by taking the first moment of $P^{\kappa}_{qq}(x)$.
The two loop splitting function for $g \rightarrow \kappa$ is evaluated
in the light-cone gauge by folding the
anomalous $- {\alpha_s \over \pi} N_f (1-x)$ factor with the splitting
function for $g \rightarrow gg$, viz. $P^g_{gg}(x)$.

Now we can deduce the general properties of the anomaly in $g_1$.
We have stressed
that the spin and anomalous components of $\Delta q(x,Q^2)$ are not
separately gauge invariant.
This means that the anomaly $\kappa$ is an intrinsic part of the
measured spin dependent quark distribution in the inclusive DIS
cross section.
The evolution equation (15a) says that the piece which is shifted
between $q_S$ and $\kappa$ scales at $O(\alpha_s)$.
We can make gauge transformations which increase the magnitude of
$\kappa$ at the expence of the spin part over all $x$.
We have shown that the anomaly scales at $O(\alpha_s)$ in
$\Delta q(x,Q^2)$ and also that it couples to the hard photon in
exactly the same way as a quark
(via the tree amplitude or $C^q(x, \alpha_s)$ in OPE language).
Hence the anomaly makes a scaling contribution to $g_1$.

The anomaly has all the characteristics of a large $x$ effect.
Certainly there is no good theoretical reason to suppose that it
is confined to small $x$
in the $g_1$ data like the ordinary gluon distribution
$\Delta g(x,Q^2)$.
A large $x$ anomaly would be seen as an OZI breaking effect in the
large $x$ bins of the polarised DIS experiment.
It would mean that the spin of the three constituent quarks is
screened by the gauge symmetry of QCD.
Since the anomaly is not present in $g_3$ it would be most helpful
to obtain data on this
C-odd spin dependent proton structure function. A large $x$ anomaly
would show up as a finite difference between $g_1$ and $g_3$ at
large $x$ [6, 7].

\section {The anomaly and the parton model: what is a parton?}

It is very important to understand how our hitherto formal discussion
of the anomaly applies to exclusive processes like jet production.
Up to this point we have been discussing the inclusive DIS cross-section
and this allowed us to use the operator product expansion.
In the OPE treatment the measured quark distribution includes
all those partons which interact locally with the hard photon (ie. they
couple to the photon via $C^q(x,\alpha_s)$).
The OPE parton distributions can be separated experimentally by taking
precise DIS data
over a range of $Q^2$ and then applying the evolution equations to fit a
phenomenological parametrisation to each of $\Delta q_0(x,Q^2)$ and
$\Delta g(x,Q^2)$.
If we wish to make predictions for jet production we normally use the
QCD improved parton model.

The improved parton model is based on the factorisation of
mass singularities.
These singularities arise when partons branch in the collinear
direction and can be
absorbed into a parton density, which is often {\it assumed} to be
equal to
the OPE parton distribution evaluated in the
light-cone gauge $A_+=0$.
The DIS cross section is then written as the convolution of these parton
densities with a ``hard" part of the phase space for photon-parton
scattering.
The improved parton model deals with explicit quark and gluon spin
degrees of freedom - it has no $\kappa$.
This conflict arises because
the improved parton model was formulated as a problem of collinear partons
whereas the anomaly is strictly a non-collinear effect.
One would never find $\kappa$ by appealing to the improved parton model
alone.
In other words, factorisation is necessary but not sufficient
to completely
define the parton model for helicity dependent processes in QCD [6].

Let us illustrate the problem by considering two-quark-jet production
in polarised photon gluon fusion.
We work in perturbation theory to $O(\alpha_s)$.
If we apply the operator product expansion to this process then the
total cross section
is written
as the sum of the OPE quark distribution $\Delta q^{(g)}(x, \mu^2)$
and the gluonic Wilson coefficient $C^g(x, {Q^2 \over \mu^2}, \alpha_s)$,
where $\mu^2 \leq Q^2$ is the renormalisation scale.
As we mentioned in Sect.3 the anomaly is included in
$\Delta q^{(g)}(x, \mu^2)$; it describes a contact interaction with the
target gluon.

In the QCD improved parton model we would say that the gluon distribution
$\Delta g(x, \mu^2)$
is characterised by
two-quark-jet events with large $k_T^2 \geq \mu^2$ [24].
On the other hand, the anomaly
corresponds to part of the phase space corresponding to two-quark-jet events
with $k_T^2 \sim Q^2$.
This contact interaction has been interpreted by some authors [20, 26]
to mean that the anomaly is a local probe of the gluon distribution
$\Delta g(x, Q^2)$ in polarised DIS.
Here we take a different view.
It is important in this example of photon gluon fusion to be precise
about when we regard the gluon as the {\it target} and when it is a
{\it parton}.
In the OPE language, the gluon should be regarded as the target when it
is used to write down the
quark distribution $\Delta q^{(g)}(x, Q^2)$ (which includes the anomaly).
It is a parton when it corresponds to the gluonic coefficient
$C^g(x, \alpha_s)$.
The large $k_T^2$ jets which are associated with the anomaly in the
photon gluon fusion process are coming from
the shattering (or jet-fragmentation) of the {\it target} gluon and we
distinguish
these jets from the jets which are associated with the
interaction of a gluon {\it parton} with the hard photon
(ie. the coefficient $C^g(x,\alpha_s)$ in the OPE, which is universal
to all targets in DIS).
This may appear semantics in the photon gluon fusion process, where
a gluon is both the target and a parton.
However, when we have a hadronic target like the proton one should
be careful about saying
that the anomaly is characterised {\it only} by
two-quark-jet events with large $k_T^2 \sim Q^2$.
We need to understand the gluonic part of the hadron wavefunction.

There is some part of the hadron wavefunction which includes
a gluon which is close to being in an asymptotic state (ie. sufficiently
off shell)
so that it may be treated perturbatively - as if we were we doing QED.
Only in this case (when $\alpha_s$ is small) can we reliably say that
the anomaly has a two quark jet signature.
(The forward matrix elements of $k_{\mu}$ are gauge invariant in QED
and the anomaly does measure the helicity of the gluon in this case.)
In general the wavefunction is non-perturbative and here one has to worry
about gauge invariance.
Since the $q_S$ and $\kappa$ components of the physical spin dependent
quark distribution are not separately gauge invariant (we consider an
arbitary gauge)
it makes no sense to associate the anomaly $\kappa$ exclusively
with two-quark jet events and
the $q_S$ parton with one jet events.
(One cannot gauge transform a one jet event into a two jet event!)

It appears that there is some critical coupling $\alpha_s^c$ such that the
anomaly is characterised by two-quark jet events for
$\alpha_s \leq \alpha_s^c$ and some (if not all) times by one jet events
for $\alpha_s \geq \alpha_s^c$.
This latter effect generates the non-perturbative anomaly contribution
to $g_A^0$, which was found by Forte [27].
(One has to be careful not to push the gauge dependence argument so far
as to claim that the non-perturbative $q_S$ and $\kappa$ components of
$\Delta q(x, Q^2)$ have the same jet signature.
This would be true if $\kappa(x,Q^2)$ could be gauge transformed to zero
in any arbitrary gauge.
There seems no reason to suppose such a strong hypothesis.)
Thus, we find that the relationship between the OPE and improved parton
model breaks down for spin dependent processes.
The pointlike interaction between the hard photon and the gluonic component
described by $\kappa(x,Q^2)$ contributes
to the semi-inclusive cross-section for one-quark jet events!
This means that one has to be careful when using perturbative QCD
to predict
absolute jet cross sections for helicity dependent processes involving
polarised initial state hadrons.

\vspace{2.0cm}
\begin{center}
{\bf Acknowledgements}
\end{center}

It is a pleasure to thank V. N. Gribov, L. Mankiewicz, N. N. Nikolaev,
A. Schafer and A. W. Thomas
for many discussions on the spin problem. \\
I am grateful to N. N. Nikolaev for helpful comments on the manuscript.

\pagebreak

\begin{center}{\bf References}\end{center}
\vspace{3ex}
\begin{enumerate}
\item
J. Ashman et al., Phys. Lett. B206 (1988) 364, Nucl. Phys. B328 (1990) 1.
\item
G. Baum et al., Phys. Rev. Lett. 51 (1983) 1135.
\item
R. Windmolders, Int. J. Mod. Phys. A7 (1992) 639.
\item
S. D. Bass and A. W. Thomas, Cavendish Preprint HEP 92/5 (1992),
to appear in J. Phys. G.
\item
G. Veneziano, Okubofest lecture, CERN preprint TH-5840/90 (1990).
\item
S. D. Bass, Zeit. Phys. C55 (1992) 653.
\item
S. D. Bass and A. W. Thomas, Cavendish preprint HEP 93/1 (1993).
\item
S. D. Bass and A. W. Thomas, Nucl. Phys. A527 (1991) 519c;
J. Ellis, M. Karliner and C. Sachrajda, Phys. Lett. B231 (1989) 497;
S. D. Bass, N. N. Nikolaev and A. W. Thomas, Adelaide preprint
ADP-90-133/T80 (1990).
\item
J. Zinn-Justin, {\it Quantum field theory and critical phenonema},
Oxford UP (1989).
\item
P. A. M. Guichon, l'Aquila lectures, Saclay preprint DPhN-Saclay-9136 (1990).
\item
J. S. Bell and R. Jackiw, Nuovo Cimento 51A (1969) 47.
\item
S. L. Adler, Phys. Rev. 177 (1969) 2426.
\item
For a review, see
R. J. Crewther, ``Chiral Properties of QCD", in Field Theoretical Methods
in Particle Physics, Kaiserslautern 1979, ed. W. Ruhl, Vol. 55B NATO
Study Institute Series, Plenum (1980) p.529.
\item
J. Kodaira, Nucl. Phys. B165 (1980) 129.
\item
W. Bardeen, Nucl. Phys. B75 (1974) 246;
R. J. Crewther, Acta Physica Austriaca Suppl. XIX 47 (1978) 47.
\item
G. 't Hooft and M. Veltman, Nucl. Phys. B44 (1972) 189.
\item
I. I. Balitsky and V. M. Braun, Phys. Lett. B267 (1991) 405.
\item
R. L. Jaffe and A. Manohar, Nucl. Phys. B337 (1990) 509.
\item
V. N. Gribov, Budapest preprint KFKI-1981-66 (1981).
\item
R. D. Carlitz, J. C. Collins and A. H. Mueller, Phys. Lett. B214 (1988) 229.
\item
V. N. Gribov, remark at SLAC Lepton Photon Symposium, see proceedings
page 59, ed. M. Riordan, World Scientific (1990).
\item
V. N. Gribov and L. N. Lipatov, Sov. J. Nucl. Phys. 15 (1972) 439;
L. N. Lipatov, Sov. J. Nucl. Phys. 20 (1974) 181;
Yu. L. Dokshitzer, Sov. Phys. JETP 46 (1977) 641;
G. Altarelli and G. Parisi, Nucl. Phys. B126 (1977) 298.
\item
D. J. Gross and F. Wilczek, Phys. Rev. D9 (1974) 980.
\item
S. D. Bass, B. L. Ioffe, N. N. Nikolaev and A. W. Thomas,
J. Moscow Phys. Soc. 1 (1991) 317.
\item
G. G. Ross and R. G. Roberts, Rutherford preprint RAL-90-062 (1990).
\item
A. V. Teryaev and O. V. Efremov, Dubna preprint E2-88-287 (1988);
G. Altarelli and G. G. Ross, Phys. Lett. B212 (1988) 391.
\item
S. Forte, Nucl. Phys. B331 (1990) 1.
\end{enumerate}

\pagebreak

\begin{center} {\bf Figures} \end{center}

\begin{enumerate}
\item
The anomaly here is the purely pointlike part of the triangle coupling
of the axial-tensor to the proton via a gluonic intermediate state.
\item
The OPE quark parton distribution of a gluon is calculated at $O(\alpha_s)$
(and in $A_+=0$ gauge) from the forward $x$ dependent triangle graph.
\end{enumerate}
\end{document}